\documentclass[runningheads]{llncs}
\usepackage{graphicx}
\usepackage{hyperref}
\usepackage{color}

\begin{document}
\title{Towards a provenance management system for astronomical observatories}
%
%
\author{
Mathieu Servillat\inst{1}\orcidID{0000-0001-5443-4128} \and
François Bonnarel\inst{2} \and
\\Catherine Boisson\inst{1}\orcidID{0000-0001-5893-1797} \and
Mireille Louys\inst{2,3}\orcidID{0000-0002-4334-1142} \and
Jose Enrique Ruiz\inst{4}\orcidID{0000-0003-3274-4445} \and
\\Michèle Sanguillon\inst{5}\orcidID{0000-0003-0196-6301}
}
\authorrunning{M. Servillat et al.}
%
\institute{
Laboratoire Univers et Théories, Observatoire de Paris, Université PSL, CNRS, Université de Paris, 92190 Meudon, France; \email{mathieu.servillat@obspm.fr} \and
Centre de Données astronomiques de Strasbourg, Observatoire Astronomique de Strasbourg, Université de Strasbourg, CNRS-UMR 7550, Strasbourg, France \and
ICube Laboratory, Université de Strasbourg, CNRS-UMR 7357, Strasbourg, France \and
Instituto de Astrofísica de Andalucía, Granada, Spain \and
Laboratoire Univers et Particules de Montpellier, Université de Montpellier, CNRS/IN2P3, France
}
\maketitle              
\begin{abstract}
We present here a provenance management system adapted to astronomical projects needs. We collected use cases from various astronomy projects and defined a data model in the ecosystem developed by the IVOA (International Virtual Observatory Alliance). From those use cases, we observed that some projects already have data collections generated and archived, from which the provenance has to be extracted (provenance “on top”), and some projects are building complex pipelines that automatically capture provenance information during the data processing (capture “inside”). Different tools and prototypes have been developed and tested to capture, store, access and visualize the provenance information, which participate to the shaping of a full provenance management system able to handle detailed provenance information.
\end{abstract}

\section{Context}

Astronomical observatories and data providers are increasingly involved in the development of Open Science. The process of making data FAIR\footnote{\url{https://www.go-fair.org/fair-principles}} (Findable, Accessible, Interoperable and Reusable) often has to be integrated early in the development of astronomical projects. Since more than 20 years, the IVOA\footnote{\url{https://www.ivoa.net}} (International Virtual Observatory Alliance) provides various standards to foster interoperability and enable the production of FAIR data.

The Reusable principle is more subjective and requires rich metadata to demonstrate the quality, reliability and trustworthiness of the data. Detailed provenance is thus a key information to provide along with the astronomical data. The IVOA validated in April 2020 a Provenance Data Model \cite{2020ivoa.spec.0411S} to structure this information. It is based on the W3C PROV concepts of Entity, Activity and Agent \cite{std:W3CProvDM} with a dedicated set of classes for activity description (e.g. method, algorithm, software) and activity configuration (e.g. parameters).

\section{Requirements and current perception of provenance}

Several use cases have been discussed within the IVOA and the European ESCAPE project \cite{B9-56_adassxxx}. Astronomical projects that produce data generally develop structured pipelines, scripts and specific methodologies to prepare data products for the end-user from raw data (acquired from observations or generated by simulation). 

Key information on what processes were applied and how they were performed is thus relevant to the end-user and could be captured directly during the process (capture “inside”). For older or other projects, a posteriori metadata extraction from data/metadata/logs (provenance “on top”) can also provide similar information, with the risk of missing details and links. We often realize too late that there are missing elements or links in the provenance, this is why the capture of the provenance should be as detailed as possible and as naive as possible (simply record what happens). In any case, the granularity of the provenance has to be adapted from one project to another.

\subsection{Basic handling of provenance}

\begin{figure}
\includegraphics[width=\textwidth]{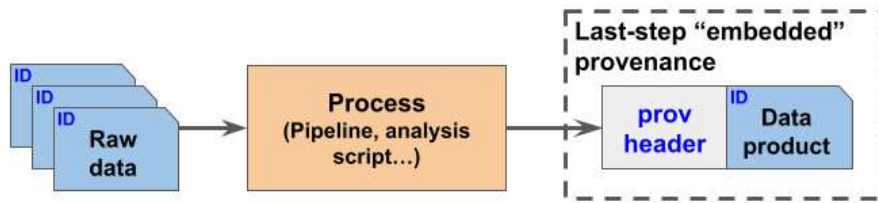}
\caption{Basic handling of provenance information.} \label{fig1}
\end{figure}

In general, the perception in the community is that provenance information is easily stored with the data, as a set of keywords recorded in the header of a data product file. This is represented in Figure~\ref{fig1}. This perception is particularly strong in astronomy with the large adoption of the FITS (Flexible Image Transport System) file format \cite{1981A&AS...44..363W}, that provides a human readable header based on keywords.

\subsection{Last-step provenance}
\label{sec:laststepprov}

The complex modeling of provenance information makes it improper to be stored as a flat list of keywords, as provenance is better represented by a graph, based on chains of activities and entities that are used and generated. We thus define the full provenance as this graph, up to the raw data, and the last-step minimum provenance as an embedded list of keywords \cite{B9-56_adassxxx}. The last-step provenance contains information on: the entity itself, one contact agent, the last activity that generated this entity. It also contains identifiers of other used and generated entities. All this information is compatible with the IVOA Provenance data model. Such a last-step provenance can thus be stored in a file header, and should moreover enable the reconstruction of the full provenance through the recursive exploration of used entities.

\section{A provenance management system}

If a basic handling of provenance information may be sufficient for some projects, it is necessary to build a more advanced provenance management system that stores this information separately, as files or in a database. Such a system is composed of the following parts :
\begin{itemize}
    \item[1/] \textbf{Capture "inside"}: provenance information is recorded during the execution of a pipeline that runs various processing steps, generates intermediate data files...
    \item[2/] \textbf{Ingestion}: the captured information is transported in a structured format that can be parsed and managed.
    \item[3/] \textbf{Storage}: the ingested information is then safely stored in a database that preserves its logic.
    \item[4/] \textbf{Visualization and exploration}: the full provenance can be queried and visualized.
\end{itemize}

\subsection{Tools, prototypes and protocols}

Several tools have been developed in relation with the IVOA Provenance data model. They are the bricks to build a full provenance management system able to handle detailed provenance information:

\begin{itemize}

    \item \texttt{voprov}\footnote{\url{https://github.com/sanguillon/voprov}}: This Python package extends the W3C PROV compatible \texttt{prov} package to implement the IVOA Provenance data model. It provides a way to create a ProvDocument object and exchange it as an XML, JSON or graphical file.

    \item \texttt{logprov}\footnote{\url{https://github.com/mservillat/logprov}}: This Python package captures provenance events when running Python functions or methods that are specifically decorated and defined. Those events are recorded through the logging system as structured dictionaries, and can then be transformed using \texttt{voprov}. This package was initially developed with the high level interface of the \texttt{gammapy} package \cite{2020ASPC..522..525L}.

    \item \textbf{ProvSAP}: a Simple Access Protocol that returns a W3C PROV file from a regular GET query on an HTTP endpoint. Arguments can be passed, such as: ID, DEPTH (ALL/1...), DIRECTION (FORWARD/BACKWARD), RESPONSEFORMAT (PROV-SVG/PROV-JSON...), MODEL (IVOA/W3C), AGENTS (0/1), CONFIGURATION (0/1), DESCRIPTIONS (0/1/2), ATTRIBUTES (0/1). This system if for example implemented in the OPUS job manager\footnote{\url{https://voparis-uws-test.obspm.fr/provsap?ID=a9b7e2}} \cite{P9-89_adassxxx} and in other tools \cite{2020ASPC..522..545S}.

    \item \textbf{ProvTAP}: IVOA Table Access Protocol using ADQL for queries and a TAP Schema, itself based on the IVOA Provenance data model \cite{2019ASPC..523..313B}. It's a reverse mechanism to locate data through queries on its provenance.
    Every feature of the model instantiated in the TAP service can then be explored. This approach enables queries to test the data quality, based on the analysis of parameters of some key activities. It is also possible to recompute datasets whose progenitors have been found erroneous. 
    
\end{itemize}

\subsection{Description of the system}

\begin{figure}
\includegraphics[width=\textwidth]{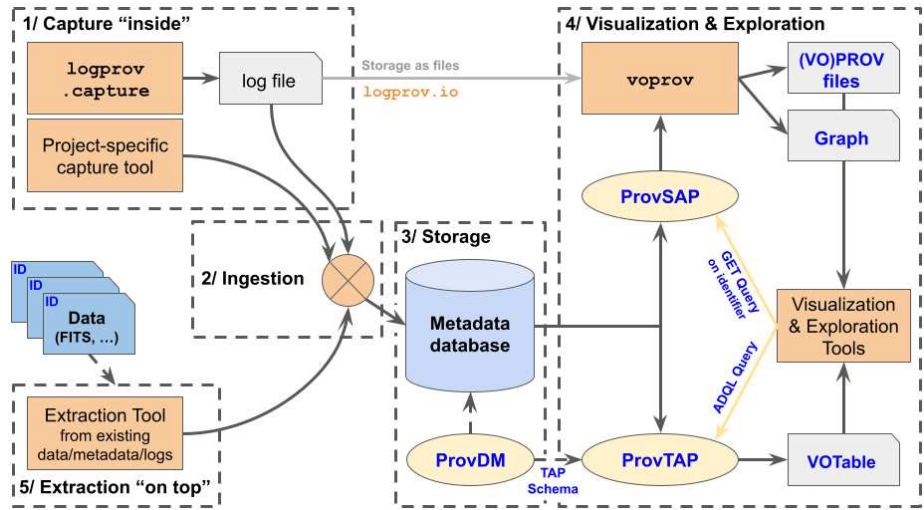}
\caption{Provenance management system.} \label{fig2}
\end{figure}

As shown in Figure~\ref{fig2}, the IVOA Provenance Data Model (ProvDM) is implemented as a relational database and connected to an access service based on the IVOA Table Access Protocol (ProvTAP) \cite{2019ASPC..523..313B}. 
A Simple Access Protocol (ProvSAP) is also being specified within the IVOA to provide directly W3C PROV files, using the \texttt{voprov} package.

In the system, provenance information is exchanged via structured logs, W3C PROV files (XML, JSON) or graphs (SVG, PNG). The \texttt{voprov} and \texttt{logprov} packages are being developed to propose a generic solution to the implementation of the system, along with project-specific capture tools (e.g. \texttt{ctapipe}\footnote{\url{https://cta-observatory.github.io/ctapipe}} or CTADIRAC\footnote{\url{https://gitlab.cta-observatory.org/cta-computing/dpps/CTADIRAC}} in the context of the Cherenkov Telescope Array \cite{P9-250_adassxxx}). 
The Visualization \& Exploration subsystem is based on standards to foster interoperability and the reuse of existing tools. 

Different implementations based on this schema are possible to adapt the provenance management to the needs and size of the project.

\subsection{Extraction "on top"}

A last block in Figure~\ref{fig2} (labelled 5/) indicates the use case of already existing data from which provenance can be extracted and ingested in the system. In many astronomy projects, some provenance information can be extracted from file headers, or from log files. Such an extraction would be more efficient if embedded provenance information were stored in a standard list of keywords such as the last-step provenance list (see \ref{sec:laststepprov}).

\section{Software and reproducibility}

Depending on the project, the workflow executed to produce science ready data (the final products) can be extracted from the provenance system designed following the IVOA strategy. For each activity execution, the input and output entities and the configuration parameters are recorded, as well as a representation of the ActivityDescription class, where the software name, version, documentation, etc, are traced.
To be fully reproducible, we envisage to access such coding blocks through the ActivityDescription class by pointing to a code repository.
This can be set up as a dictionary of codes within a specific project, as in the CTA pipeline or other under development projects such as Euclid, LSST, etc. 

Software can also be shared within the community and curated in code registries, such as the Software Heritage \cite{dicosmo:hal-01590958}, or the astronomy dedicated software published in ASCL\footnote{\url{http://ascl.net}} (Astrophysics Source Code Library), 
or for multi-messenger astronomy, the future ESCAPE OSSR project\footnote{\url{https://wiki.escape2020.de/index.php/WP3_-_OSSR}}.

Many astronomical projects deal with large amounts of data and require increasing computation power. This has pushed forward the development of science platforms that implement the code-to-the-data strategy. In this new computing and distributing architecture, rich metadata profiles to describe the provenance of datasets and the code applied to process them, is a key for reproducibility and interoperability.

\subsection*{Acknowledgements}
We acknowledge support from the ESCAPE project funded by the EU Horizon 2020 research and innovation program (Grant Agreement n°824064). Additional funding was provided by the INSU (Action Sp\'ecifique Observatoire Virtuel, ASOV), the Action F\'ed\'eratrice CTA at the Observatoire de Paris and the Paris Astronomical Data Centre (PADC).

%
%
%
\bibliographystyle{splncs04}
\bibliography{Servillat_AstroProv}
%




\end{document}